\documentclass[epj,nopacs]{svjour}
\usepackage{graphics}
\def\lsim{\mathrel{\raise.3ex\hbox{$<$\kern-.75em\lower1ex\hbox{$\sim$}}}}
\def\gsim{\mathrel{\raise.3ex\hbox{$>$\kern-.75em\lower1ex\hbox{$\sim$}}}}
\begin{document}
%
%
\title{Little Higgs Phenomenology}
\author{Heather E. Logan}
\institute{Department of Physics, University of Wisconsin, Madison, Wisconsin
53706, USA}
\date{Received: date / Revised version: date}
%
\abstract{
Recently a new class of models has emerged that addresses the naturalness
problem of a light Higgs boson.  In these ``little Higgs'' models, the
Standard Model Higgs boson is a pseudo-Nambu-Goldstone boson of an
approximate global symmetry.
The Higgs boson acquires mass radiatively only through
``collective breaking'' of the global symmetry, so that more than
one interaction is required to give the Higgs a mass.
This protects the Higgs mass from receiving quadratically divergent
radiative corrections at one-loop.
These models contain new vector bosons, fermions and scalars
at the TeV scale that cancel the
quadratic divergences in the Higgs mass due to the Standard Model
gauge, top quark, and Higgs boson loops.  In this talk I review the
phenomenology of the little Higgs models, focusing on collider signatures
and electroweak precision constraints.
} 
\maketitle

\section{Introduction}
\label{intro}

The Standard Model (SM) of strong and electroweak interactions has
passed stringent tests up to the highest energies accessible today.
The precision electroweak data \cite{Hagiwara:fs} point to the existence of
a light Higgs boson in the SM, with mass $m_H \lsim 200$ GeV.
The SM with such a light Higgs boson can be viewed as an effective
theory valid up to a much higher energy scale $\Lambda$, 
which could be as high as the Planck scale.
In particular, the precision electroweak data exclude
dimension-six operators arising from strongly coupled new physics
below a scale $\Lambda$ of order 10 TeV \cite{Barbieri};
any new physics appearing below this scale must be weakly coupled.
However, without protection by a symmetry, the Higgs mass is quadratically
sensitive to the cutoff scale $\Lambda$ via quantum corrections,
rendering the theory with $m_H \ll \Lambda$ rather unnatural.
For example, for $\Lambda = 10$ TeV, the ``bare'' Higgs mass-squared
parameter must be tuned against the quadratically divergent radiative
corrections at the 1\% level.
This gap between the electroweak scale $m_H$ and the cutoff 
$\Lambda$ is called the ``little hierarchy''.

Little Higgs models
\cite{decon,minmoose,Littlest,SU6Sp6,KaplanSchmaltz,custodialmoose,SkibaTerning,Spencer}
revive an old idea to keep the
Higgs boson naturally light: they make the Higgs particle a
pseudo-Nambu-Goldstone boson \cite{PNGBhiggs} of a broken global symmetry.
The new ingredient of little Higgs models is that at least two
interactions are needed to explicitly break all of the global symmetry
that protects the Higgs mass.  This forbids quadratic divergences in the
Higgs mass at one-loop; the Higgs mass is then smaller than the cutoff
scale $\Lambda$ by {\it two} loop factors, making the cutoff scale
$\Lambda \sim 10$ TeV natural and solving the little hierarchy problem.

From the bottom-up point of view, the most important quadratic divergences
in the Higgs mass due to top quark, gauge boson, and Higgs boson loops
are canceled by loops of new weakly-coupled
fermions, gauge bosons, and scalars with masses around a TeV.  In
contrast to supersymmetry, the cancellations in little Higgs models
occur between loops of particles with the {\it same} statistics.
Electroweak symmetry breaking is triggered by a
Coleman-Weinberg \cite{coleman-weinberg} potential generated by
integrating out the heavy degrees of freedom.

The constraints on little Higgs models from electroweak precision data
have been examined in detail in
Refs.~\cite{GrahamEW1,JoAnneEW,GrahamEW2,JayEW}.
The constraints come from $Z$ pole data from LEP and SLD,
low-energy neutrino-nucleon scattering, atomic parity violation,
and the $W$ boson mass measurement from LEP-II and the Tevatron.
These measurements probe contributions from the exchange of virtual
heavy gauge bosons between fermion pairs, the modification of $Z$-pole
observables due to mixing of the $Z$ with the heavy gauge bosons,
and the shift in the mass ratio of the $W$ and $Z$.
The lower bounds on the masses of the new heavy gauge bosons are
generally in the 1.5--2 TeV range \cite{GrahamEW2,JayEW}.
The electroweak precision measurements
tend to favor parameter regions in which the new
heavy gauge bosons are approximately decoupled from the SM fermions,
thereby suppressing four-fermi interactions.
The electroweak precision measurements do not directly constrain the mass
of the top-partner.  However, the mass of the top-partner is related to
the heavy gauge boson masses by the structure of the model.  For naturalness,
the top-partner should be as light as possible.  The lower bounds on the
top-partner mass are generally in the 1--2 TeV range.

The ``Littlest Higgs'' model \cite{Littlest} is a minimal model of this type.
It consists of a nonlinear sigma model with a global
SU(5) symmetry which is broken down to SO(5) by a vacuum condensate
$f \sim \Lambda/4\pi \sim$ TeV.
The gauged subgroup [SU(2)$\times$U(1)]$^2$ is broken at the
same time to its diagonal subgroup SU(2)$\times$U(1), identified
as the SM electroweak gauge group.
The breaking of the global symmetry leads to 14 Goldstone bosons, four of
which are eaten by the broken gauge generators, leading to four
massive vector bosons: an SU(2) triplet $Z_H$, $W^{\pm}_H$, and a
U(1) boson $A_H$.
The ten remaining uneaten Goldstone bosons
transform under the SM gauge group as a doublet $h$ (which becomes
the SM Higgs doublet) and a triplet $\phi$ (which gets a mass of order $f$).
A vector-like pair of colored Weyl fermions is also needed to cancel
the divergence from the top quark loop, leading to a new heavy vector-like
quark with charge $+2/3$.  In this talk I review the phenomenology
of the Littlest Higgs model, following Refs.~\cite{LHpheno,LHloop}.

\section{Collider phenomenology}
\label{sec:pheno}

\begin{figure}
\resizebox{0.50\textwidth}{!}{\includegraphics{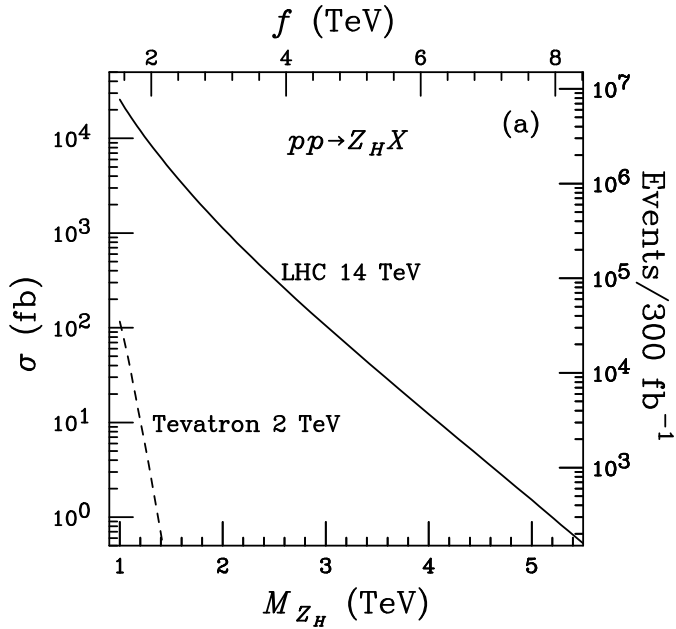}}
\caption{Cross section for $Z_H$ production in Drell-Yan at the LHC
and Tevatron, for $\cot\theta = 1$.  From \cite{LHpheno}.}
\label{fig:sigmaZH}
\end{figure}

The heavy SU(2) gauge bosons $Z_H$ and $W_H$ can be produced via Drell-Yan
at the LHC (and at the Tevatron, if they are light enough).
The cross section is proportional to $\cot^2\theta$ because the
$Z_H$ and $W_H$ couplings to fermion pairs are proportional to 
$\cot\theta \equiv g_2/g_1$ (see Ref.~\cite{LHpheno}).
In Fig.~\ref{fig:sigmaZH} we show the cross section for $Z_H$ production
at the Tevatron and LHC for $\cot\theta = 1$.
In the region of small $\cot\theta \simeq 0.2$ favored 
\cite{GrahamEW1,JoAnneEW,GrahamEW2,JayEW}
by the precision
electroweak data, the cross section must be scaled down by
$\cot^2\theta \simeq 0.04$.  Even with this suppression factor,
a cross section of 40 fb is expected at the LHC for $M_{Z_H} \simeq 2$ TeV,
leading to 4,000 events in 100 fb$^{-1}$ of data.
The production and decay of $Z_H$ and $W_H$ at the LHC has also been studied
in Ref.~\cite{gustavo}.

The $Z_H$ boson decays to fermion pairs with partial widths proportional to 
$\cot^2\theta$ and to boson pairs ($ZH$ and $W^+W^-$) with partial widths
proportional to $\cot^2 2\theta$.
This feature can be used to distinguish the Littlest
Higgs model from a ``big Higgs'' model with the same gauge group in which
the Higgs doublet transforms under only one of the SU(2) groups
\cite{gustavo}, in which case the $ZH$ and $W^+W^-$ partial widths
would be proportional to $\cot^2\theta$.
Neglecting final-state masses, the branching fraction into three
flavors of charged leptons is equal to that into one flavor of quark
($\simeq 1/8$ for $\cot\theta \gsim 0.5$), due
to the equal coupling of $Z_H$ to all SU(2) fermion doublets.  The branching
ratio into $ZH$ is equal to that into $W^+W^-$.
The decay branching fractions of $W_H$ follow a similar pattern.

The Littlest Higgs model also contains a heavy U(1) gauge boson, $A_H$,
which is generally the lightest new particle in the model.
Its couplings to fermions are more model dependent than
those of $Z_H$ and $W_H$, since they depend on the U(1) charges
of the fermions (see Ref.~\cite{LHpheno} for details).
Even the presence of $A_H$ is somewhat model-dependent, since
this particle can be eliminated by gauging only
one U(1) group (hypercharge) without adding a significant amount of
fine-tuning \cite{GrahamEW2}.

\begin{figure}
\resizebox{0.5\textwidth}{!}{\includegraphics{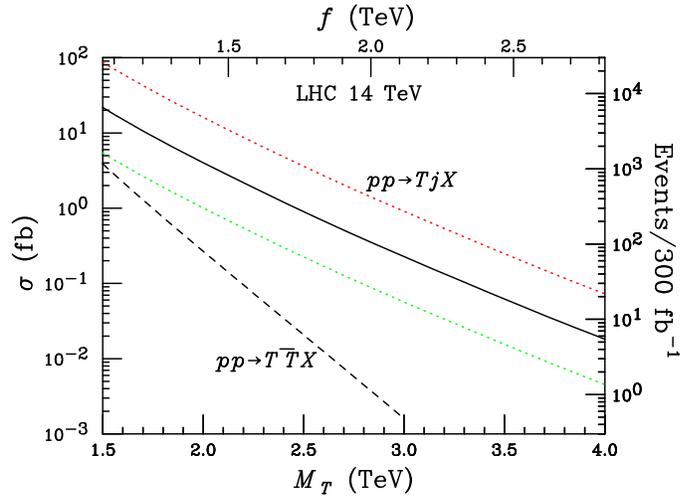}}
\caption{Cross sections for $T$ production at the LHC.  The single $T$
cross section is shown for $\lambda_1/\lambda_2 = 1$ (solid) and
$\lambda_1/\lambda_2 = 2$ (upper dotted) and $1/2$ (lower dotted).
The QCD pair production cross section is shown for comparison (dashed).
From \cite{LHpheno}.}
\label{fig:toph}
\end{figure}

The heavy top-partner $T$ can be pair-produced via QCD with
model-independent couplings.  The cross section for this production mode 
falls quickly with increasing $M_T$ due to phase space suppression.
The single $T$ production mode, $W^+b \to T$, dominates for $M_T \gsim$ TeV
(Fig.~\ref{fig:toph}).  
The cross section for single $T$ production depends on the ratio of couplings
$\lambda_1/\lambda_2$ (see Ref.~\cite{LHpheno}), which relates $M_T$ 
to the scale $f$.
$T$ decays into $tH$, $tZ$, and $bW^+$ with branching fractions
1/4, 1/4, and 1/2, respectively.
The top sector is quite similar in many of the other little Higgs models
in the literature, so these general features of $T$ production and
decay should apply.  Some models contain more than one
top-partner \cite{SU6Sp6,KaplanSchmaltz,custodialmoose,Spencer,Anntalk} or
contain partners for all three fermion generations 
\cite{KaplanSchmaltz,SkibaTerning}; in these cases the phenomenology 
will be modified.

The decay partial widths of the Higgs boson into gluon pairs or photon
pairs are modified in the Littlest Higgs model by the new heavy particles
running in the loop and by the shifts in the Higgs couplings
to the SM $W$ boson and top quark \cite{LHloop}.
These modifications of the Higgs couplings to gluon or photon pairs
scale like $1/f^2$, and thus decouple at high $f$ scales.
The range of partial widths for given $f$ values accessible
by varying the other model parameters was computed in Ref.~\cite{LHloop}.

For $f \gsim 1$ TeV, the correction to $\Gamma(H \to gg)$ is unlikely
to be observable because of the large SM QCD uncertainty \cite{QCDggh}.
The correction to $\Gamma(H \to \gamma\gamma)$ is more promising;
it could be observed at a photon collider,
where the $\gamma\gamma \to H \to b \bar b$ rate can be measured
to about 2\% \cite{gammagamma} for $m_H \sim$ 115--120 GeV.
Combining this with BR($H \to b \bar b$) measured to about 1.5--2\% 
at an $e^+e^-$ collider \cite{e+e-}
allows the extraction of $\Gamma(H \to \gamma\gamma)$ with a precision
of about 3\%.
Such a measurement would be sensitive to $f < 2.7$ (1.8, 1.2) TeV at the 
$1\sigma$ ($2\sigma$, $5\sigma$) level.
For comparison, the electroweak precision constraints require
$f \gsim 1$ TeV in the Littlest Higgs model \cite{GrahamEW2}.

\section{Conclusions}
\label{sec:concl}
The little Higgs idea provides a new way to address the little hierarchy
problem of the Standard Model by making the Higgs a pseudo-Nambu-Goldstone
boson of a spontaneously broken global symmetry.  The global symmetry
is explicitly broken by gauge and Yukawa interactions; however, no
single interaction breaks all the symmetry protecting the Higgs mass.
This prevents quadratically divergent radiative corrections to the Higgs
mass from appearing at the one-loop level, and thus allows the
cutoff scale to be pushed higher by one loop factor, to $\sim 10$ TeV.
From the bottom-up point of view, the quadratically divergent radiative
corrections to the Higgs mass due to top quark, gauge boson, and Higgs
loops are canceled by new heavy quarks, gauge bosons, and scalars,
respectively.

The details of the phenomenology depend on the specific model.
Since quite a few little Higgs models have appeared 
over the past two years, finding generic features of the
phenomenology is important.
Very generically, there must be new gauge bosons, fermions and scalars
to cancel the quadratic divergences in the Higgs mass.

There is some tension between the precision electroweak constraints
pushing up the new particle masses and the requirement that the new
particles be light to avoid fine tuning.
However, by tuning the parameters of the models appropriately one can
satisfy both constraints.  This tuning of the parameters should be
explained in the ultraviolet completion of the nonlinear sigma model.
Our developing understanding of the effects of
little Higgs models on the electroweak precision observables is now
driving model building to incorporate features that loosen the constraints.
Taking these constraints into account,
the new particles should live in the 1--2 TeV mass range
and should be accessible at the LHC.

\begin{acknowledgement}
It is a pleasure to thank Tao Han, Bob McElrath and Lian-Tao Wang
for collaborations leading to the papers \cite{LHpheno,LHloop} on
which this talk is based.  I also thank the organizers of EPS 2003
for the opportunity to speak.
This work was supported by the U.S.~Department of Energy
under grant DE-FG02-95ER40896
and by the Wisconsin Alumni Research Foundation.
\end{acknowledgement}


\end{document}